\documentclass[10pt,a4paper]{article}
\usepackage{bm, amssymb,pifont,cancel, amsmath,comment,color,authblk}
\usepackage[dvips]{graphicx}
\usepackage{multicol}
\newcommand{\cW}{\mathcal{W}}
\newcommand{\cF}{\mathcal{F}}
\begin{document}
\begin{flushright}
WU-HEP-15-05\\
KEK-TH-1801
\end{flushright}
\begin{center}
{\Large{\bf{Matter coupled Dirac-Born-Infeld action \\in 4-dimensional ${\cal N}=1$ conformal supergravity}}}\\ 
\vskip 6pt
\large{Hiroyuki Abe}${}^1${\renewcommand{\thefootnote}{\fnsymbol{footnote}}\footnote[1]{E-mail address: abe@waseda.jp}}, \large{Yutaka Sakamura}${}^{2,3}${\renewcommand{\thefootnote}{\fnsymbol{footnote}}\footnote[2]{E-mail address: sakamura@post.kek.jp}} and \large{Yusuke Yamada}${}^1${\renewcommand{\thefootnote}{\fnsymbol{footnote}}\footnote[3]{E-mail address: yuusuke-yamada@asagi.waseda.jp}}\\
\vskip 4pt
${}^1${\small{\it Department of Physics, Waseda University,}}\\
{\small{\it Tokyo 169-8555, Japan}}\\
\vskip 1.0em

${}^2${\small\it KEK Theory Center, Institute of Particle and Nuclear Studies, 
KEK, \\ Tsukuba, Ibaraki 305-0801, Japan} \\ \vspace{1mm}
${}^3${\small\it Department of Particles and Nuclear Physics, \\
SOKENDAI (The Graduate University for Advanced Studies), \\
Tsukuba, Ibaraki 305-0801, Japan} 

\end{center}
\begin{abstract}
We construct the Dirac-Born-Infeld action in the context of ${\cal N}=1$ conformal supergravity and its possible extensions including matter couplings. We especially focus on the Volkov-Akulov constraint, which is important to avoid ghost modes from the higher derivative terms. In the case with matter couplings, we find the modified D-term potential. 
\end{abstract}
\begin{multicols}{2}
\section{Introduction}
The Dirac-Born-Infeld (DBI) action~\cite{Born:1934gh,Dirac:1962iy} was constructed as a generalization of the Maxwell action including higher derivative correction, in a way that no ghost modes appear. The DBI action is also considered as the action which describes the dynamics of D-branes in superstring theory~\cite{Bergshoeff:1996tu,Aganagic:1996nn,Tseytlin:1999dj}. From a phenomenological and theoretical viewpoint, it is interesting to construct the supersymmetric (SUSY) version of the DBI action. Such a generalization has been studied in global SUSY~\cite{Cecotti:1986gb,Bagger:1996wp,Rocek:1997hi,Ketov:1998ku} and in supergravity (SUGRA)~\cite{Cecotti:1986gb,Kuzenko:2002vk,Kuzenko:2005wh,Brinne:1999vv}.\footnote{See Refs.~\cite{Tseytlin:1999dj,Ketov:2001dq} for reviews.} Recently, the DBI action in $\cal{N}$-extended SUSY in various dimensions were developed in Refs.~\cite{Bergshoeff:2013pia,Bellucci:2013mha}.

The DBI action has a scale which will be related to a fundamental scale, such as the Planck scale or the string scale. If that scale is close to the Planck scale, we cannot neglect the gravitational effects and have to embed it into SUGRA. There are several works along this direction. The authors of Ref.~\cite{Cecotti:1986gb} suggest a way to the SUGRA extensions of the DBI action. In Ref.~\cite{Kuzenko:2002vk}, the action in Ref.~\cite{Bagger:1996wp} is extended to SUGRA, and its component action is also studied by the same authors~\cite{Kuzenko:2005wh}. 
However, the matter coupled DBI action in SUGRA has been less studied. For example, in Ref.~\cite{Kuzenko:2005wh}, it is discussed for a special case. We will clarify how matter multiplets can couple to the DBI sector in more general cases. 

In this work, we will work in the superconformal formulation of 4-dimensional ${\cal N}=1$ SUGRA~\cite{Kaku:1978nz,Kugo:1982cu},
 which has the larger symmetry than the Poincar\'e SUSY. This formulation is useful thanks to the larger symmetry, especially to construct the matter coupled SUGRA. In ${\cal N}=1$  global SUSY, the authors of Ref.~\cite{Bagger:1996wp} found that the SUSY DBI action can be constructed from two chiral multiplets related to each other by a certain condition. We propose the corresponding condition in conformal SUGRA, and use it as a guiding principle to construct the matter coupled DBI action. We will also emphasize that the condition implies the Volkov-Akulov constraint~\cite{Volkov:1972jx,Rocek:1978nb}, which is important to avoid ghost modes from higher derivative terms.

The remaining parts of this paper are organized as follows. In Sec.~\ref{global BI}, we briefly review the construction of the DBI action in global SUSY case, and show how the Volkov-Akulov constraint works. Sec.~\ref{SUGRA} is devoted to embed the DBI action into conformal SUGRA. We first generalize the condition in global SUSY to the one in conformal SUGRA, and construct the DBI action minimally coupled to gravity. In Sec.~\ref{modification}, we discuss the matter coupled version of the DBI action and find the modified D-term potential including all the higher order corrections, which has not been known so far. We will give comments on the relation between our results and the action derived in Refs.~\cite{Cecotti:1986gb,Kuzenko:2002vk,Kuzenko:2005wh} in Sec.~\ref{comments}. We also discuss the relation between our model and the DBI action in superstring theory. Finally, we give a summary and discuss the possible extensions in Sec.~\ref{summary}. 
\section{Dirac-Born-Infeld action in global supersymmetry}\label{global BI}
We briefly review the construction of SUSY DBI action proposed by J. Bagger and A. Galperin~\cite{Bagger:1996wp}. Let us consider two ${\cal N}=1$  chiral superfields $X$ and ${\cW}_\alpha$, where $\alpha$ denotes the spinor index, ${\cW}_\alpha=-\frac{1}{4}\bar{D}^2D_\alpha V$and $V$ is an abelian gauge vector superfield in ${\cal N}=1$  SUSY and $D_\alpha$ and $\bar{D}_{\dot{\alpha}}$ are an ${\cal N}=1$  SUSY covariant derivative and its complex conjugate respectively. To construct the SUSY DBI action, we impose the following condition\footnote{In Ref.~\cite{Bagger:1996wp}, the authors derived the condition~(\ref{BG cond}) from the non-linear realization of 4D ${\cal N}=2$ SUSY. However we just use the condition as a guideline to construct the DBI action in this work . We will briefly comment on the relation between our construction and partial SUSY breaking in Sec.~\ref{summary}.} on $X$ and ${\cW}_\alpha$,
\begin{align}
X=\cW^\alpha\cW_\alpha+\frac{1}{4}X\bar{D}^2\bar{X},\label{BG cond}
\end{align}
where $\bar{X}$ is the complex conjugate of $X$. One can solve Eq.~(\ref{BG cond}) algebraically, and obtain 
\begin{align}
X=\cW^2+\frac{1}{2}\bar{D}^2\left[ \frac{\cW^2\bar{\cW}^2}{1-\frac{1}{2}+\sqrt{1-A+B^2/4}}\right],\label{BG sol}
\end{align}
where 
\begin{align}
A\equiv \frac{1}{2}(D^2\cW^2+\bar{D}^2\bar{\cW}^2),\\
B\equiv \frac{1}{2}(D^2\cW^2-\bar{D}^2\bar{\cW}^2).
\end{align}
By using $X=X(\cW,\bar{\cW})$, we can describe the supersymmetric DBI action as,
\begin{align}
\mathcal{L}_{DBI}=\frac{1}{4}\int d^2\theta X(\cW,\bar{\cW})+{\rm h.c.}.\label{SUSY BI}
\end{align}
The bosonic part of the action~(\ref{SUSY BI}) is 
\begin{align}
\mathcal{L}_{DBI}|_{B}&=1-\Biggl( 1+\frac{1}{2}\cF_{ab}\cF^{ab}+\frac{1}{8}(\cF_{ab}\cF^{ab})^2\nonumber\\
&\quad \quad-\frac{1}{4}\cF_{ab}\cF^{bc}\cF_{cd}\cF^{da}\Biggr)^{1/2}\nonumber\\
&=1-\sqrt{-{\rm det}(\eta_{ab}+\cF_{ab})},\label{flat BI}
\end{align}
where $\cdot|_B$ denotes the bosonic part of $\cdot$, and $\eta_{ab}$ is the Minkowski metric.

From Eq.~(\ref{BG sol}), we find that $X^2(\cW,\bar{\cW})=0$ because $X(\cW,\bar{\cW})$ is proportional to $\cW^2$ and Grassmannian property of $\cW_\alpha$. This feature shows the underlying Volkov-Akulov supersymmetry, that is, the non-linearly realized SUSY.  

We can also express the action~(\ref{SUSY BI}) with Lagrange multiplier multiplets as follows,  
\begin{align}
\mathcal{L}=& \int d^2\theta \left[X+\Lambda(\cW^2+\frac{1}{4}X\bar{D}^2\bar{X}-X)+MX^2\right]+{\rm h.c.},\label{multiplier}
\end{align} 
where $\Lambda$ and $M$ are Lagrange multiplier chiral multiplets. Note that $X$ in Eq.~(\ref{multiplier}) is an unconstrained chiral multiplet. The equations of motion of $\Lambda$ and $M$ give the constraint~(\ref{BG cond}) and $X^2=0$ respectively. The latter condition $X^2=0$ becomes trivial after solving the constraint~(\ref{BG cond}), therefore the term $MX^2$ in Eq.~(\ref{multiplier}) does not change the resultant action after solving the first condition~(\ref{BG cond}) and we obtain the action~(\ref{SUSY BI}). However, the last term in Eq.~(\ref{multiplier}) is useful to represent the underlying Volkov-Akulov constraint on $X$. 

Before embedding the DBI action into conformal SUGRA, we mention why the Volkov-Akulov constraint is important. By using the superspace identity~$-1/4\int d^2\theta \bar{D}^2(\cdots)=\int d^4\theta (\cdots)+{\rm tot.div}$, we can rewrite the action in Eq.~(\ref{multiplier}) as 
\begin{align}
\mathcal{L}=&-\int d^4\theta(\Lambda+\bar{\Lambda})|X|^2\nonumber\\
&+\left( \int d^2\theta \left[X+\Lambda(\cW^2-X)+MX^2\right]+{\rm h.c.}\right).\label{multiplier2}
\end{align}
The first term in Eq.~(\ref{multiplier2}) corresponds to the K\"ahler potential~$K=-(\Lambda+\bar{\Lambda})|X|^2$, which gives kinetic terms of the scalar components of $\Lambda$ and $X$. The kinetic coefficients are given by the K\"ahler metric~$K_{I\bar{J}}=\partial_I\partial_{\bar{J}}K$, and in this case, the determinant of the K\"ahler metric becomes $-|\hat{X}|^2$ where $\hat{X}$ is the scalar component of $X$. This negative definite determinant means that there is at least one ghost mode among the scalar components of $\Lambda$ and $X$. However, due to the Volkov-Akulov constraint~$X^2=0$, $\hat{X}$ is equivalent to a fermion bilinear $\sim \psi^X\psi^X/F^X$. Therefore, the bosonic part of the determinant vanishes, and then the scalar ghost mode disappears. Therefore, the underlying Volkov-Akulov constraint~$X^2=0$ is important for avoiding ghost modes. Indeed, the DBI action does not contain any bosonic ghosts, and so we consider the generalization of the DBI action in which the Volkov-Akulov constraint is satisfied. 

It is worth noting that the Volkov-Akulov type supersymmetry always appear in DBI type action as discussed, e.g., in Refs.\cite{Aganagic:1996nn,Bergshoeff:2013pia,Kallosh:1997aw}. It is also found that such a nonlinear SUSY can play important roles in cosmology, particle phenomenology and moduli stabilization in superstring theory~\cite{Antoniadis:2014oya,Ferrara:2014kva,Aoki:2014pna,Dall'Agata:2014oka,Kallosh:2014hxa}. The DBI type action of anti D3 brane, which can realize de Sitter vacua in superstring theory, was also discussed in Refs.~\cite{Kallosh:2014wsa,Bergshoeff:2015jxa}.
\section{Dirac-Born-Infeld action in 4D ${\cal N}=1$  conformal supergravity}\label{SUGRA}
In this section, we generalize the SUSY DBI action discussed in Sec.~\ref{global BI} to the one in ${\cal N}=1$  conformal SUGRA.\footnote{In the following discussions, we use the notation in Ref.~ \cite{Kugo:1982cu}, but the F-term density formula $[\cdots]_F$ in this paper should be understood as $[\cdots]_F+{\rm h.c.}$ in Ref.~\cite{Kugo:1982cu}.} The generalization is useful for construction of the matter coupled DBI action, which will be discussed in Sec.~\ref{modification}. 

In conformal SUGRA, supermultiplets are characterized by the weights $(w,n)$ for the dilatation and  the U$(1)_A$ symmetry\footnote{U$(1)_A$ is an automorphism of the 4 dimensional ${\cal N}=1$  superconformal algebra.} called the Weyl and the chiral weights respectively. For example, gauge vector multiplets should have $(w,n)=(0,0)$. For chiral multiplets, their weights should satisfy the condition $w=n$, but the value of $w(=n)$ can be arbitrary. The weights of field strength superfields are determined as $(w,n)=(3/2,3/2)$. 

Let us generalize the condition~(\ref{BG cond}) to the one in conformal SUGRA. We assume that the weights of $X$ and $\bar{X}$ are $(w,w)$ and $(w,-w)$ respectively. The first term on the right hand side of Eq.~(\ref{BG cond}) has $(w,n)=(3,3)$. All of the terms in Eq.~(\ref{BG cond}) are chiral multiplets, and therefore the term $X\bar{D}^2\bar{X}$ should be replaced with $X\Sigma(\bar{X})$, where $\Sigma$ is the chiral projection operator in conformal SUGRA. However, the chiral projection can be applied to the multiplets satisfying $w-n=2$, then the Weyl weight of $X$ is determined as $w=1$. Therefore, the term $X\Sigma(\bar{X})$ has the weights $(w,n)=(3,3)$ which is the same with those of $\cW^2$, because the chiral projection operator has the weights $(w,n)=(1,3)$. On the other hand, the term on the left hand side of Eq.~(\ref{BG cond}) has $(w,n)=(1,1)$, which conflicts with the ones of the right hand side. This implies that we have to impose the so-called chiral compensator multiplet $S_0$ with $(w,n)=(1,1)$.\footnote{We can also use other compensator multiplets, such as a real and a complex linear compensators. The expressions of Eq.~(\ref{SC BG cond}) with different compensators are summarized in Appendix.~\ref{appB}. }
By introducing $S_0$, we can write the naive extension of Eq.~(\ref{BG cond}) as follows,
\begin{align}
S_0^2X=\cW^2-aX\Sigma(\bar{X}), \label{preSC BG cond}
\end{align}
where $a$ is a complex constant, but it can be real by field redefinitions without loss of generality, so that we take it as a real parameter. For simplicity of the following discussions, we replace $X$ with $S_0X$, then the weights of $X$ are $(w,n)=(0,0)$ and we can rewrite Eq.~(\ref{preSC BG cond}) as 
\begin{align}
S_0^3X=\cW^2-aS_0X\Sigma(\bar{S}_0\bar{X}).\label{SC BG cond}
\end{align}

Formally Eq.~(\ref{SC BG cond}) can be rewritten as 
\begin{align}
X=\frac{\cW^2}{S_0(S_0^2+a\Sigma(\bar{S}_0\bar{X}))}.
\end{align}
As in the global SUSY case, $X$ is proportional to $\cW^2$, and that implies $X^2=0$ even in conformal SUGRA. Therefore, the following action can be considered as the conformal SUGRA generalization of SUSY DBI action~(\ref{multiplier}),
\begin{align}
S=&[bS_0^3X]_F+[2\Lambda(\cW^2-aS_0X\Sigma(\bar{S}_0\bar{X})-S_0^3X)]_F\nonumber\\
&+[S_0^3MX^2]_F+[-c|S_0|^2]_D,\label{SC BI action}
\end{align}
where $[\cdots]_{F,D}$ denotes the F- and D-term density formulae for chiral multiplets with $(w,n)=(3,3)$ and for real multiplets with $(w,n)=(2,0)$ respectively, $b,c$ are real constants, and we have chosen the weights of $\Lambda$ and $M$ as $(w,n)=(0,0)$. We multiplied the second term in Eq.~(\ref{SC BI action}) by $2$ to simplify the following analysis. We added the last term in order to obtain the Einstein-Hilbert term.

We use the action~(\ref{SC BI action}) instead of directly solving (\ref{SC BG cond}). In the following discussion, we only focus on the bosonic part of the action~(\ref{SC BI action}). First, we use the equation of motion of $M$, which leads $X^2=0$. Then, the scalar component $\hat{X}$ can be written by its fermionic partner $\psi^X$ and F-term~$F^X$ as $\hat{X}\sim(\psi^X\psi^X)/F^X$ as in the case of global SUSY. This feature incredibly reduces the bosonic part of Eq.(\ref{SC BI action}) because we can ignore the terms containing $\hat{X}$. 

We apply the following theorem shown in Ref.~\cite{Cecotti:1987sa}, $[-2\Lambda S_0X\Sigma(\bar{S}_0\bar{X})]_F=[|S_0|^2(\Lambda+\bar{\Lambda})|X|^2]_D+{\rm tot.div}$, and obtain the bosonic part of the action~(\ref{SC BI action}) as
\begin{align}
S|_B=&\int d^4x\sqrt{-g}\Biggl[2a|S_0|^2(\Lambda+\bar{\Lambda})|F^X|^2\nonumber\\
&+\{S_0^3(b-2\Lambda)F^X+{\rm h.c.}\}+2({\rm Re}\Lambda)\cF^{\mu\nu}\cF_{\mu\nu}\nonumber\\
&-2i({\rm Im}\Lambda)\cF_{\mu\nu}\tilde{\cF}^{\mu\nu}-4({\rm Re}\Lambda)D^2\nonumber\\
&+\frac{c}{3}|S_0|^2R-c(|F^{S_0}|^2-D_\mu S_0D^\mu\bar{S}_0)\Biggr],\label{comp1}
\end{align}
where we represent the scalar component of each supermultiplet by the same letter as the supermultiplet itself, $D$ is the D-term of the vector multiplet, $\mu,\nu$ denote the curved spacetime indices, and $\tilde{\cF}^{\mu\nu}\equiv -\frac{i}{2}\epsilon^{\mu\nu\rho\sigma}\cF_{\rho\sigma}$, $D_\mu S_0=(\partial_\mu-b_\mu-A_\mu)S_0$, $b_\mu$ and $A_\mu$ are the gauge fields for dilatation and U$(1)_A$ symmetry respectively, $F^{S_0}$ denotes the F-term of $S_0$. Note that $c$ should be positive to obtain the correct kinetic term of the graviton.    

To obtain the action in Einstein frame, we set the following superconformal gauge fixing conditions~\cite{Kugo:1982cu,Kugo:1982mr}, 
\begin{align}
S_0|_B=\bar{S}_0|_B=\sqrt{\frac{3}{2c}}, \quad b_\mu=0,\label{pure gauge}
\end{align}
where the first condition is for the dilatation and U$(1)_A$ symmetry and the second one for the special conformal symmetry, and we use the Planck unit convention $M_{pl}(=2.4\times 10^{18}{\rm GeV}) =1$.  For simplicity, we assume $c=3/2$ in the following, then $S_0=\bar{S}_0=1$. 

After the superconformal gauge fixings, the action~(\ref{comp1}) becomes
\begin{align}
S|_B=&\int d^4x\sqrt{-g}\Biggl[2a(\Lambda+\bar{\Lambda})|F^X|^2\nonumber\\
&+\{(b-2\Lambda)F^X+{\rm h.c.}\}+2({\rm Re}\Lambda)\cF^{\mu\nu}\cF_{\mu\nu}\nonumber\\
&-2i({\rm Im}\Lambda)\cF_{\mu\nu}\tilde{\cF}^{\mu\nu}-4({\rm Re}\Lambda)D^2\nonumber\\
&+\frac{1}{2} R-\frac{3}{2} (|F^{S_0}|^2-A^\mu A_\mu)\Biggr].\label{comp3}
\end{align}
We can easily eliminate $A_\mu$, F-, and D-terms from Eq.(\ref{comp3}), and obtain the following on-shell action,
\begin{align}
S|_B^{\mbox{\scriptsize on-shell}}=&\int d^4x\sqrt{-g}\Biggl[\frac{1}{2}R+2\lambda\cF_{\mu\nu}\cF^{\mu\nu}\nonumber\\
&-2i\chi\cF_{\mu\nu}\tilde{\cF}^{\mu\nu}+\frac{(b-2\lambda)^2+4\chi^2}{4a\lambda}\Biggr],\label{comp4}
\end{align}
where $\lambda\equiv{\rm Re}\Lambda$ and $\chi\equiv{\rm Im}\Lambda$.

Finally, we obtain the following conditions from variations of the auxiliary fields $\lambda$ and $\chi$,
\begin{align}
\frac{\chi}{\lambda}&=ai\cF_{\mu\nu}\tilde{\cF}^{\mu\nu},\label{c1}\\
\frac{b^2}{4\lambda^2}&=1+2a\cF_{\mu\nu}\cF^{\mu\nu}+a^2(\cF_{\mu\nu}\tilde{\cF}^{\mu\nu})^2.\label{c2}
\end{align}
By substituting the conditions (\ref{c1}) and (\ref{c2}) to the action~(\ref{comp4}), we derive the final form of the action,
\begin{align}
S|_B^{\mbox{\scriptsize on-shell}}&=\int d^4x\sqrt{-g}\Biggl[\frac{1}{2}R-\frac{b}{a}\nonumber\\
&+\frac{b}{a}\sqrt{1+2a\cF_{\mu\nu}\cF^{\mu\nu}+a^2(\cF_{\mu\nu}\tilde{\cF}^{\mu\nu})^2}\Biggr]\nonumber\\
&=\int d^4x\sqrt{-g}\left[\frac{1}{2}R-\frac{b}{a}\right]\nonumber\\
&+\frac{b}{a}\int d^4x\sqrt{-{\rm det}(g_{\mu\nu}+2\sqrt{a}\cF_{\mu\nu})}.\label{comp fin}
\end{align}
Assuming $a=1/4$ and $b=-1/4$, this action~(\ref{comp fin}) is the supergravity extension of the DBI action~(\ref{flat BI}), which contains not only the DBI term but also the kinetic term of the graviton. We can also derive the ferminonic part of the SUGRA DBI action in the same way. \footnote{However, due to the Volkov-Akulov constraint $\hat{X}\sim \psi^X\psi^X/F^X$, the elimination of $F^X$ is expected to be very complicated, so the derivation of the full-action requires further investigations. }
\section{Modification of supergravity Dirac-Born-Infeld action}\label{modification}
In this section, we discuss the possible modifications of the SUGRA DBI action~(\ref{comp fin}), which respects the Volkov-Akulov constraint~$X^2=0$. In Sec.~\ref{SUGRA}, we have discussed the case in which there are only the gravity and the vector multiplet. Here, let us consider a case that there are matter chiral multiplets that couple to the vector multiplet. 

In Sec.~\ref{SUGRA}, we have derived the condition~(\ref{SC BG cond}) as a superconformal version of Eq.~(\ref{BG cond}). We can generalize the condition~(\ref{SC BG cond}) preserving the Volkov-Akulov constraint $X^2=0$ as
\begin{align}
S_0^3X=\cW^2-S_0X\Sigma(\omega(\Phi^I,\bar{\Phi}^{\bar{J}})\bar{S}_0\bar{X}),\label{mod BG}
\end{align}
where $\omega(\Phi^I,\bar{\Phi}^{\bar{J}})$ is an arbitrary real function of matter multiplets $\Phi^I$ and $\bar{\Phi}^{\bar{J}}$. Instead of solving Eq.~(\ref{mod BG}), let us consider the following action,
\begin{align}
S=&\left[\frac{1}{2}S_0\bar{S}_0\Omega(\Phi^I,\bar{\Phi}^{\bar{J}})\right]_D+[S_0^3f(\Phi^I)X]_F\nonumber\\
&+[2\Lambda(\cW^2-S_0X\Sigma(\omega(\Phi^I,\bar{\Phi}^{\bar{J}})\bar{S}_0\bar{X})-S_0^3X)]_F\nonumber\\
&+[S_0^3MX^2]_F +[S_0^3W(\Phi^I)]_F\label{mS1}
\end{align}
where $\Omega(\Phi^I,\bar{\Phi}^{\bar{J}})$ is a real function of matters, which is related to the physical K\"ahler potential $K$ through $K=-3\log (-\Omega/3)$, and $f(\Phi^I)$ and $W(\Phi^I)$ denote arbitrary holomorphic functions of $\Phi^I$. Comparing the action~(\ref{SC BI action}) and (\ref{mS1}), we find that the generalizations correspond to the following replacement of the parameters in Eq.~(\ref{SC BI action}),
\begin{align}
a\to \omega(\Phi^I,\bar{\Phi}^{\bar{J}}), \quad b\to f(\Phi^I),\quad c\to \Omega(\Phi^I,\bar{\Phi}^{\bar{J}}).
\end{align}

The bosonic part of Eq.~(\ref{mS1}) is obtained as follows.
\begin{align}
S|_B=&\int d^4x\sqrt{-g}\Biggl[-\frac{1}{6}|S_0|^2\Omega R+2\omega|S_0|^2(\Lambda+\bar{\Lambda})|F^X|^2\nonumber\\
&+\{ S_0^3 (f-2\Lambda)F^X+{\rm h.c.}\}+2({\rm Re}\Lambda)\cF_{\mu\nu}\cF^{\mu\nu}\nonumber\\
&-2i({\rm Im}\Lambda)\cF_{\mu\nu}\tilde{F}^{\mu\nu}-4({\rm Re}\Lambda)D^2\nonumber\\
&-i|S_0|^2\Omega_Ik^ID+\mathcal{L}_{\rm ordinary}\Biggr]\label{mS2}
\end{align}
where $\Omega_I=\partial \Omega/\partial \Phi^I$, $k^I$ is a Killing vector for U(1) isometry on the manifold spanned by $\Phi^I$, and $\mathcal{L}_{\rm ordinary}$ is the parts of matter action which takes an ordinary form of the SUGRA action (see Eq.~(\ref{ordinary})). To obtain the Einstein frame action, we put the following superconformal gauge fixing conditions~\cite{Kugo:1982mr},
\begin{align}
S_0=\bar{S}_0=\sqrt{-\frac{3}{\Omega}},\quad b_\mu=0.\label{scgauge}
\end{align}
We also integrate out the auxiliary fields, such as F- and D-terms, and then obtain the following on-shell action,
\begin{align}
S|_B^{\mbox{\scriptsize on-shell}}=&\int d^4x\sqrt{-g}\Biggl[\frac{1}{2}R+2\lambda \cF_{\mu\nu}\cF^{\mu\nu}-2i\chi \cF_{\mu\nu}\tilde{\cF}^{\mu\nu}\nonumber\\
&+\frac{e^{2K/3}|f-2\Lambda|^2}{4\omega \lambda}-\frac{(K_Ik^I)^2}{16\lambda}+\mathcal{L}_{\rm ordinary}^{\mbox{\scriptsize on-shell}}\Biggr]\label{mS3}
\end{align}
where we have used the relation $\Omega=-3e^{-K/3}$, $K_I=\partial_IK$, $\lambda={\rm Re}\Lambda$, and $\chi={\rm Im}\Lambda$. $\mathcal{L}^{\mbox{\scriptsize on-shell}}_{\rm ordinary}$ is obtained as follows,
\begin{align}
\mathcal{L}_{\rm ordinary}^{\mbox{\scriptsize on-shell}}=&-K_{I\bar{J}}{\cal D}_\mu\Phi^I{\cal D}^{\mu}\bar{\Phi}^{\bar{J}}\nonumber\\
&-e^K\left(K^{I\bar{J}}D_IWD_{\bar{J}}\bar{W}-3|W|^2\right),\label{ordinary}
\end{align}
where ${\cal D}_\mu\Phi^I=\partial_\mu\Phi^I-v_\mu k^I$, $v_\mu$ is the gauge field for the U(1) symmetry, and $D_IW=\partial_IW+K_IW$.

Finally, we have to eliminate the Lagrange multipliers $\lambda$ and $\chi$. We summarize the detailed calculations for solving the equations of motion of $\lambda$ and $\chi$ in Appendix.~\ref{app}. By using the results, we obtain the following action,
\begin{align}
S|_B^{\mbox{\scriptsize on-shell}}=&\int d^4x\sqrt{-g}\left[\frac{1}{2}R-\frac{e^{\frac{2K}{3}}{\rm Re}f}{\omega}+\mathcal{L}_{\rm ordinary}^{\mbox{\scriptsize on-shell}}\right]\nonumber\\
+&\int d^4x\frac{e^{\frac{2K}{3}}{\rm Re}f\sqrt{P}}{\omega}\sqrt{-{\rm det}(g_{\mu\nu}+2e^{-\frac{K}{3}}\sqrt{\omega}\cF_{\mu\nu})}\nonumber\\
-&\int d^4x\sqrt{-g}i{\rm Im}f\cF_{\mu\nu}\tilde{\cF}^{\mu\nu},\label{mS4}
\end{align}
where
\begin{align}
P\equiv 1-\frac{\omega e^{-2K/3}(K_Ik^I)^2}{4({\rm Re}f)^2}.
\end{align}
The action (\ref{mS4}) is the generalized DBI action including matter couplings. 

For simplicity, we assume $k^I=0$, then the action~(\ref{mS4}) can be expanded with respect to $\omega$ as
\begin{align}   
S|_B^{\mbox{\scriptsize on-shell}}\to&\int d^4x\sqrt{-g}\left[-i{\rm Im}f\cF_{\mu\nu}\cF^{\mu\nu}+{\rm Re}f\cF_{\mu\nu}\cF^{\mu\nu}\right]\nonumber\\
&+\mathcal{O}(\omega)+\cdots,\label{mS5}
\end{align}
where ellipsis denotes terms irrelevant to the DBI action. The first line of Eq.~(\ref{mS5}) represents an ordinary vector action with a gauge kinetic function $f(\Phi^I)$ in ${\cal N}=1$  SUGRA. Thus, the choice of $\omega$ only affects the higher order corrections. 

Let us discuss the D-term potential in our case. By setting $\cF_{\mu\nu}=0$, the first and the second lines of Eq.~(\ref{mS4}) lead to the potential,
\begin{align}
V_D=\frac{e^{2K/3}{\rm Re}f}{\omega}\left(1-\sqrt{1-\frac{e^{-2K/3}\omega(K_Ik^I)^2}{4({\rm Re}f)^2}}\right).\label{VDK}
\end{align}
We can expand $V_D$ in terms of $\omega$, and obtain   
\begin{align}
V_D=\frac{1}{8({\rm Re}f)}(K_Ik^I)^2+\mathcal{O}(\omega).\label{VDA}
\end{align}
This is the usual D-term potential in SUGRA. 
The D-term potential~(\ref{VDK}) includes all of the higher order corrections induced by the higher derivative terms of the vector multiplet. This modified D-term potential may be interesting from the viewpoint of the D-term inflation models in SUGRA, and we will discuss it in the subsequent work~\cite{Abe:2015fha}.
\section{Some comments}\label{comments}
\subsection{Relation between our results and other works}
In this subsection, we clarify our results and the ones in Refs.~\cite{Cecotti:1986gb,Kuzenko:2002vk,Kuzenko:2005wh}.

The authors of Ref.~\cite{Cecotti:1986gb} showed the DBI action which is not coupled to matters through the supergravitational effects, equivalently, the action should take the following form,
\begin{align}
S|_B^{\mbox{\scriptsize on-shell}}=&\int d^4x\sqrt{-g}\left[\frac{1}{2}R+1+\mathcal{L}_{\rm ordinary}^{\mbox{\scriptsize on-shell}}\right]\nonumber\\
-&\int d^4x\sqrt{-{\rm det}(g_{\mu\nu}+\cF_{\mu\nu})}.
\end{align}
As you can easily find, this action can be reproduced by choosing $k^I=0$, $\omega=e^{2K/3}/4$ and $f=-1/4$ in the action~(\ref{mS4}). We clarify the reason why the gravitational couplings between the vector multiplet and matters disappear with the choice $\omega=e^{2K/3}/4$. Let us return to the constraint~(\ref{mod BG}). Remembering the relation $\Omega=-3e^{-K/3}$, the constraint~(\ref{mod BG}) with $\omega=e^{2K/3}/4$ becomes
\begin{align}
S_0^3X&=\cW^2-\frac{9}{4}S_0X\Sigma(\Omega^{-2}\bar{S}_0\bar{X})\nonumber\\
&=\cW^2-\frac{9}{4}\Sigma(\Omega^{-2}|X|^2|S_0|^2),\label{Kinv1}
\end{align}
where we have used an identity $S_0X\Sigma(\cdots)=\Sigma(S_0X \cdots)$. With the redefinition of $S_0^3X\to \tilde{X}$, we can rewrite the constraint~(\ref{Kinv1}) as
\begin{align}
\tilde{X}=\cW^2-\frac{9}{4}\Sigma\left(\frac{|\tilde{X}|^2}{(|S_0|^2\Omega)^2}\right).
\end{align}
The combination $|S_0|^2\Omega$ becomes constant after superconfomal gauge fixings with Eq.~(\ref{scgauge}), and therefore, the resulting DBI action does not involve the matter couplings through the gravitational effects. This type of $\omega$ corresponds to the coupling referred to as the K\"ahler covariant coupling in Ref.~\cite{Cecotti:1986gb}.

The procedure in Refs.~\cite{Kuzenko:2002vk,Kuzenko:2005wh} can be can be regarded as follows. First, they solved the constraint~(\ref{app3}), and obtain the DBI action with a real linear compensator, which corresponds to the DBI action in the new minimal SUGRA. Then, they introduced the matter couplings which possess the non-linear self duality of the DBI sector. The matter coupling extension can be regarded as the special choice of $\tilde\omega$ in Eq.~(\ref{app4}). They also added the usual matters to the matter coupled DBI action in the new minimal SUGRA, and they derived the old minimal version of it by using the linear-chiral duality. As a result, the action becomes a special case of our model~(\ref{mS4}). In fact, with $K=K(\Phi^i,\bar{\Phi}^{\bar{j}})-\log (i(\Phi-\bar{\Phi})/2)$, $\omega=-\kappa({\rm Im}\Phi)e^{-2K/3}$, $f=i\Phi/4$, and $\Phi=a-ie^{-\varphi}$, we can reproduce the action in Eq~(4.27) in Ref.~\cite{Kuzenko:2005wh}.
\subsection{String theoretical aspects}\label{string}
In this section, we discuss the relation between the DBI action derived in this work and the D-brane action in superstring theory.

The D3-brane action is given by 
\begin{align}
S_{\rm D3}=\int d^4\sigma\sqrt{-{\rm det}(\hat{g}_{\mu\nu}+\cF_{\mu\nu}+B_{\mu\nu})}\label{D3action}
\end{align}
where $\sigma^\mu$ denotes the world volume coordinate, $\hat{g}_{\mu\nu}$ is the induced metric on the world volume of the D3-brane, and $B_{\mu\nu}$ is an antisymmetric tensor called the B-field. The induced metric is expressed as $\hat{g}_{\mu\nu}=\partial_\mu X^M(\sigma)g_{MN}\partial_\nu X^N(\sigma)$ where $g_{MN}$ $(M,N=0,\cdots,9)$ is the 10 dimensional metric, and $X^M(\sigma)$ is a world volume scalar field.

Comparing Eq.~(\ref{D3action}) with Eq.~(\ref{comp fin}), we find that the action in Eq.~(\ref{comp fin}) may be understood as the 4D effective D3 brane action in the static gauge~$\partial_\mu X^M=\delta_\mu^M$, which leads to $g_{\mu\nu}=\hat{g}_{\mu\nu}$, under the assumption that the moduli and B-fields are fixed to 0.  

We also considered the case that matter multiplets are directly coupled to the DBI action. Such a situation occurs when the U(1) gauge symmetry is anomalous. In string theory, such an anomaly must be cancelled by the so-called Green-Schwarz (GS) mechanism~\cite{Green:1984sg,Lopes Cardoso:1991zt}. If the GS mechanism works, it is known that the gauge kinetic function of the corresponding vector multiplet includes the GS multiplet, which is charged under the vector multiplet and appears in the D-term potential. Our result~(\ref{VDK}) is consistent with this observation since the leading term~(\ref{VDA}) agrees with the known results.
\section{Summary and discussions}\label{summary}
In this work, we have constructed the generalized DBI action based on the condition derived in Ref.~\cite{Bagger:1996wp}. We respect the Volkov-Akulov constraint $X^2=0$ to avoid the higher derivative ghosts. 

In Sec.~\ref{SUGRA}, we have discussed the superconformal extension of Eq.~(\ref{BG cond}), and derived a SUGRA version of DBI action minimally coupled to gravity. In the case with matter couplings discussed in Sec.~\ref{modification}, we have generalized the condition~(\ref{SC BG cond}) to (\ref{mod BG}), which also preserves the Volkov-Akulov constraint $X^2=0$. The resultant action becomes a non-trivial form given in Eq.~(\ref{mS4}). We have shown that the action~(\ref{mS4}) becomes the ordinary vector multiplet action at the leading order. We have derived the D-term potential, which has nontrivial higher order couplings in matter fields. It is interesting to investigate how the corrections affect high energy physics, such as inflation. We will discuss this issue in the subsequent work~\cite{Abe:2015fha}. 

We have also discussed the relation between the generalized DBI action and the D-brane action in superstring theory. We have interpreted our action~(\ref{mS4}) as the D3-brane action in the static gauge and in the absence of the moduli and the B-fields. We do have to know all the couplings including these fields. To investigate such couplings, it is important to extend our construction to one in higher dimensional SUGRA. In Refs.~\cite{Bagger:1996wp,Rocek:1997hi}, the authors obtained the condition~(\ref{BG cond}) from the partial breaking of 4D ${\cal N}=2$ SUSY to ${\cal N}=1$. Therefore, we expect that it is possible to construct the DBI action for Dp-branes from the partial breaking of higher dimensional SUGRA. That will be also our future investigation.

Finally, let us comment on the relation between our construction and partial ${\cal N}=2$ SUSY breaking discussed in Refs.\cite{Bagger:1996wp,Rocek:1997hi}. The authors of Refs.\cite{Bagger:1996wp,Rocek:1997hi} discussed the global SUSY case. In such a case, there is a Goldstino multiplet for the broken SUSY. In SUGRA, it is expected that the Goldstino multiplet is absorbed and forms an ${\cal N}=1$ massive spin $3/2$ multiplet combined with a part of the ${\cal N}=2$ gravitational multiplet. If we consider our action as the low-energy effective action for partial SUSY breaking, such a massive multiplet is already integrated out. Thus, in the case that our Maxwell multiplet $\cW_{\alpha}$ is the Goldstino multiplet just like in Ref.~\cite{Bagger:1996wp}, it should be unphysical and is subject to the constraint~(\ref{SC BG cond}) whose form is determined by the ${\cal N}=2$ SUSY algebra. On the other hand, in the case that it is a physical massless multiplet, it is no longer the Goldstino multiplet so that the constraint it satisfies can be released from (\ref{SC BG cond}) and take more general form~(\ref{mod BG}). In both cases, the Volkov-Akulov constraint should be satisfied in order to avoid ghost modes.
\section*{Acknowledgments}
H.A., Y.S. and Y.Y. are supported in part by Grant-in-Aid for Young Scientists (B) 
(No. 25800158),
Grant-in-Aid for Scientific Research (C) (No.25400283),    
and Research Fellowships for Young Scientists (No.26-4236), 
which are from Japan Society for the Promotion of Science, respectively.  
\begin{appendix}
\section{The constraints~(\ref{SC BG cond}) and (\ref{mod BG}) with different compensators}\label{appB}
As shown in Ref.~\cite{Kugo:1982cu}, there are three types of irreducible compensator multiplets, which give different sets of auxiliary fields in the gravity multiplet. One is the chiral compensator we discuss in this paper, and the others are a real and a complex linear multiplets denoted by $L$ and ${\bf L}$ respectively. The Weyl and the chiral weights of $L$ and ${\bf L}$ are $(2,0)$ and $(w,w-2)$ respectively. 

By using the identity $S_0X\Sigma(\bar{S}_0\bar{X})=\Sigma(|S_0X|^2)$, Eq.~(\ref{SC BG cond}) can be rewritten as follows,
\begin{align}
S_0^3X=\cW^2-a\Sigma(|S_0X|^2).\label{app1}
\end{align}
In the case that a compensator is $L$ or ${\bf L}$, the weights of the multiplet on the left hand side of Eq.(\ref{app1}) can not be compensated, because the compensator $L$ (or ${\bf L}$) is not a chiral multiplet but the both sides of Eq.(\ref{app1}) should consist of chiral multiplets. Therefore, $X$ should be a chiral multiplet with $(w,n)=(3,3)$ in a such case. Then the expression~(\ref{app1}) should be replaced with 
\begin{align}
X=\cW^2-a\Sigma(|X|^2|S_0|^{-4}).\label{app2}
\end{align}
In Eq.~(\ref{app2}), the compensator is not required to be a chiral multiplet. Therefore, we can write down the alternate conditions of Eq.~(\ref{SC BG cond}) with $L$ or ${\bf L}$ as,
\begin{align}
X=& \cW^2-a\Sigma(|X|^2L^{-2}),\label{app3}\\ 
X=& \cW^2-a\Sigma(|X|^2|{\bf L}|^{-4/w}),
\end{align}
where $w$ is the Weyl weight of ${\bf L}$.

As in the same way, we can obtain the alternate conditions of Eq.~(\ref{mod BG}),
\begin{align}
X=&\cW^2-a\Sigma (\tilde{\omega}(L,\Phi^I,\bar{\Phi}^{\bar{J}})|X|^2L^{-2}),\label{app4}\\
X=&\cW^2-a\Sigma (\hat{\omega}({\bf L},,\Phi^I,\bar{\Phi}^{\bar{J}})|X|^2|{\bf L}|^{-4/w}),
\end{align}
where $\tilde{\omega}(L,\Phi^I,\bar{\Phi}^{\bar{J}})$ and $\hat{\omega}({\bf L},,\Phi^I,\bar{\Phi}^{\bar{J}})$ are real functions of matters and $L$ or ${\bf L}$ and $\bar{\bf L}$, with $(w,n)=(0,0)$.
\section{Integrating out the Lagrange multipliers}\label{app}
We show the detailed calculations for deriving the action~(\ref{mS4}) from (\ref{mS3}) in Sec.~\ref{modification}. The relevant Lagrangian terms for integrating out the multipliers in Eq.~(\ref{mS3}) are  as follows,
\begin{align}
\mathcal{L}_{\lambda,\chi}=&2\lambda\cF_{\mu\nu}\cF^{\mu\nu}-2i\chi\cF_{\mu\nu}\tilde{\cF}^{\mu\nu}\nonumber\\
&+\frac{[(p-2\lambda)^2+(q-2\chi)^2]}{4\omega e^{-2K/3}\lambda}-\frac{(K_Ik^I)^2}{16\lambda}\label{ap1},
\end{align}
where $\lambda={\rm Re}\Lambda$, $\chi={\rm Im}\Lambda$, $p={\rm Re}f(\Phi)$, and $q={\rm Im}f(\Phi)$.

By varying $\lambda$ and $\chi$ in Eq.(\ref{ap1}), we obtain the following equations,
\begin{align}
&2\lambda^2\cF_{\mu\nu}\cF^{\mu\nu}+\frac{(K_Ik^I)^2}{16}-\frac{[(p-2\lambda)^2+(q-2\chi)^2]}{4\omega e^{-2K/3}}\nonumber\\
&+\frac{\lambda(2\lambda-p)}{\omega e^{-2K/3}}=0,\label{ap2}\\
&2\chi-q=2i\omega e^{-2K/3}\lambda\cF_{\mu\nu}\tilde{\cF}^{\mu\nu}.\label{ap3}
\end{align}
By substituting Eq.~(\ref{ap3}) into Eq.~(\ref{ap2}), we obtain
\begin{align}
&\left[2\cF_{\mu\nu}\cF^{\mu\nu}+\frac{1}{\omega e^{-2K/3}}+\omega e^{-2K/3}(\cF_{\mu\nu}\tilde{\cF}^{\mu\nu})^2\right]\lambda^2\nonumber\\
&=\frac{p^2}{4\omega e^{-2K/3}}-\frac{(K_Ik^I)^2}{16}.\label{ap4}
\end{align}
From Eqs.~(\ref{ap3}) and (\ref{ap4}), we derive the following field values of $\lambda$ and $\chi$,
\begin{align}
\lambda=&\pm\frac{p}{2\sqrt{\omega}e^{-K/3}}\left(1-\frac{\omega e^{-2K/3}(K_Ik^I)^2}{4p^2}\right)^{\frac{1}{2}}\nonumber\\
&\times\left(2\cF_{\mu\nu}\cF^{\mu\nu}+\frac{1}{\omega e^{-\frac{2K}{3}}}+\omega e^{-\frac{2K}{3}}(\cF_{\mu\nu}\tilde{\cF}^{\mu\nu})^2\right)^{-\frac{1}{2}},\label{ap5}\\
\chi=&\frac{q}{2}+i\omega e^{-2K/3}\lambda\cF_{\mu\nu}\tilde{\cF}^{\mu\nu}.\label{ap6}
\end{align}
$\lambda$ has two solutions with opposite signs, however, by requiring that the action vanishes when $k^I=\cF_{\mu\nu}=0$, the field value is uniquely determined as the one with a positive sign. Using the solutions~(\ref{ap5}) and (\ref{ap6}), we can obtain the final result~(\ref{mS4}) after some calculations.
\end{appendix}

\end{multicols}
\end{document}